\documentclass[aps,prl,twocolumn,longbibliography,notitlepage]{revtex4-1}

\usepackage{epsfig,amssymb,amsmath,amsthm,amsfonts,amsbsy,mathrsfs}

\begin{document}
\title{Aging Maxwell fluids}
\author{Jie Lin}
\affiliation{Center for Quantitative Biology and Peking-Tsinghua Center for Life Sciences, Academy for Advanced Interdisciplinary Studies, Peking University, Beijing 100871, China}
\date{\today}
\begin{abstract}
	Many experiments show that protein condensates formed by liquid-liquid phase separation exhibit aging rheological properties. Quantitatively, recent experiments by Jawerth et al. [Science 370, 1317, 2020] show that protein condensates behave as aging Maxwell fluids with an increasing relaxation time as the condensates age. Despite the universality of this aging phenomenon, a theoretical understanding of this aging behavior is lacking. In this work, we propose a mesoscopic model of protein condensates in which a phase transition from aging phase to non-aging phase occurs as the control parameter changes, such as temperature. The model predicts that protein condensates behave as viscoelastic Maxwell fluids at all ages, with the macroscopic viscosity increasing over time. The model also predicts that protein condensates are non-Newtonian fluids under a constant shear rate with the shear stress increasing over time. Our model successfully explains multiple existing experimental observations and also makes general predictions that are experimentally testable.
\end{abstract}

\maketitle
Aging phenomena are widely observed in various soft matter systems, including polymers \cite{Hutchinson1995}, soft glasses \cite{Cipelletti2003, Bonn2017}, disordered mechanical systems \cite{Lahini2017}. Recently, many experiments found that protein condensates formed by liquid-liquid phase separation \cite{Brangwynne2009, Brangwynne2015, Mao2019}, both in vivo and in vitro, also exhibit aging behavior: their dynamics slow over time and become more solid-like \cite{Patel2015, Lin2015, Woodruff2017, Shin2017, Banani2017, Franzmann2018, Wang2018, Berry2018}. More quantitatively, Jawerth et al. \cite{Jawerth2020} studied the aging of protein condensates by investigating their rheological properties in the linear viscoelasticity regime using laser tweezers \cite{Jawerth2018}. The authors found that the complex moduli of protein condensates are self-similar at all ages --- they exhibit the same viscoelastic behavior of Maxwell fluids with a relaxation time increasing as the material ages. Meanwhile, the material appeared amorphous at all times. This aging behavior appears generic because it is observed in various protein condensates, including PGL-3 and FUS family proteins. Previous models on the aging rheology of disordered materials mostly discussed yield stress materials -- finite stress is required to shear the material in the zero shear rate limit \cite{Sollich1997, Sollich1998, Fielding2000, Derec2001, Fielding2009, Sollich2017, Parley2020}. For example, the soft glassy rheology model \cite{Sollich1997, Sollich1998, Fielding2000} shows that a glassy material behaves as a Maxwell fluid at high temperature but without aging; aging occurs at low temperature, but the material also becomes solid. As far as we know, the mechanism of the aging Maxwell fluid behavior of protein condensates is far from clear despite its universality and fundamental importance to cell biology and disease \cite{Shin2017, Jawerth2020}.

In this manuscript, we propose a mesoscopic model of protein condensates in which the condensate is considered as a combination of many mesoscopic parts. The system's macroscopic rheological response is the average of all the mesoscopic parts. In the following, we first introduce the model in the linear viscoelasticity regime in which a phase transition from aging phase to non-aging phase occurs when the temperature is above some critical value. We then study the aging phase and show that protein condensates behave as aging Maxwell fluids with increasing characteristic relaxation times as the materials age. We then switch to nonlinear rheology and show that for aging Maxwell fluids, the shear stress increases with time under constant shear rate -- protein condensates are rheopectic fluids. Finally, we show that the stress-strain curves of aging Maxwell fluids under different shear rates are invariant if the products of shear rate and characteristic relaxation time are the same.

\begin{figure}[htb!]
	\includegraphics[width=0.48\textwidth]{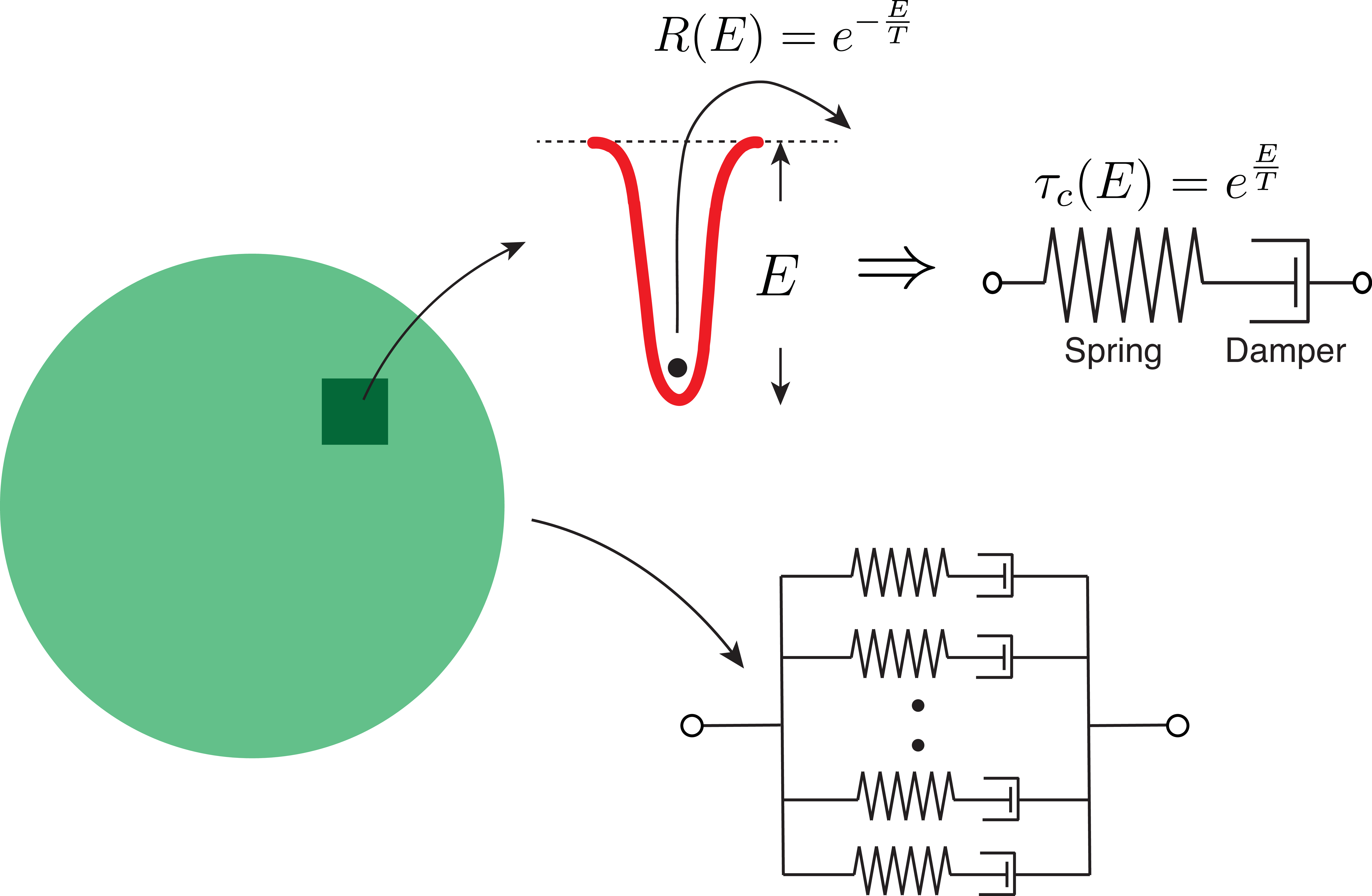}
	\caption{A protein condensate is conceptually subdivided into mesoscopic elements. Each element is in a metastable state with yield energy $E$, and the hopping rate to escape the metastable state is $R(E)=e^{-E/T}$ where $T$ is the temperature. Each element behaves as a mesoscopic Maxwell fluid in the linear viscoelasticity regime with the relaxation time set by its current yield energy. The whole system's rheological response is a parallel combination of all the mesoscopic elements. }\label{model}
\end{figure}

{\it Aging Maxwell fluid model ---} A protein condensate, which can be formed by liquid-liquid phase separation, is conceptually subdivided into mesoscopic parts, called elements in the following. Their sizes are small enough so that a macroscopic droplet contains a large number of them. Meanwhile their sizes are large enough so that their deformations can be described by an elastic strain variable. Because condensate-forming proteins have large regions of disorder, they can have a large number of molecular configurations \cite{Banani2017}. Therefore, each element can transit between a large number of metastable states \cite{Chebaro2015}. We assume that each element is in one metastable state with a yield energy $E$, which is the energy barrier the element needs to overcome to escape the metastable state. Therefore, the hopping rate for the element to hop out of the metastable state is $R(E)=\Gamma_0 \exp(-E/T)$. Here $\Gamma_0$ is the attempt frequency for hops, $T$ is the temperature, and the Boltzmann constant is set as $1$. In the following, we set $\Gamma_0^{-1}$ as the time unit. The model is summarized in Figure \ref{model}.

In the regime of linear viscoelasticity, the local stress of elements with yield energy $E$ evolves as $d\langle \sigma(E)\rangle/dt=k \dot{\gamma} - \exp(-E/T)\langle \sigma(E)\rangle$. Here $\gamma$ is the shear strain, ${\dot{\gamma}}$ is the shear rate, and $k$ is the elastic modulus. The bracket represents an average over an ensemble of elements with the same yield energy. In the following, we set $k$ as the stress unit. Note that in principle, the hopping rate should be $\exp(-(E-l^2/2)/T)$ where $l$ is the local strain of one element, but since we consider the linear viscoelasticity, the higher-order terms are neglected \cite{Sollich1998}. Note that the speedup of hopping rate by external shear will be important as we later discuss nonlinear rheology.

The whole system can be considered as a combination of a large number of mesoscopic Maxwell fluids connected in parallel. For a local element with yield energy $E$, the Fourier transformation of its stress in the frequency space follows $\sigma(E,\omega)=\frac{i \omega \tau_c(E)}{1+i \omega \tau_c(E)}\gamma(\omega)$ where $\tau_c(E)=\exp(\beta E)$ and $\beta=1/T$. Since the macroscopic shear stress is the average over the local stresses of all elements, the whole system's complex modulus, defined as $\sigma(\omega)=G(\omega)\gamma(\omega)$, becomes
\begin{equation}
G(\omega)=\Big\langle \frac{i \omega \tau_c(E)}{1+i\omega\tau_c(E)}\Big\rangle_E.\label{G}
\end{equation}
The average is over the yield energy distribution $P(E,t)$, which can depend on time $t$. We note that rigorously speaking, the above equation requires that $P(E,t)$ changes little in the duration of the rheology measurement so that a quasi-equilibrium yield energy distribution is a good approximation. We will show later that current experiments support that this quasi-equilibrium condition is generally satisfied. Therefore, to find the complex modulus, we only need to find the yield energy distribution $P(E,t)$, which we discuss in the following.

\begin{figure}[b!]
	\includegraphics[width=0.4\textwidth]{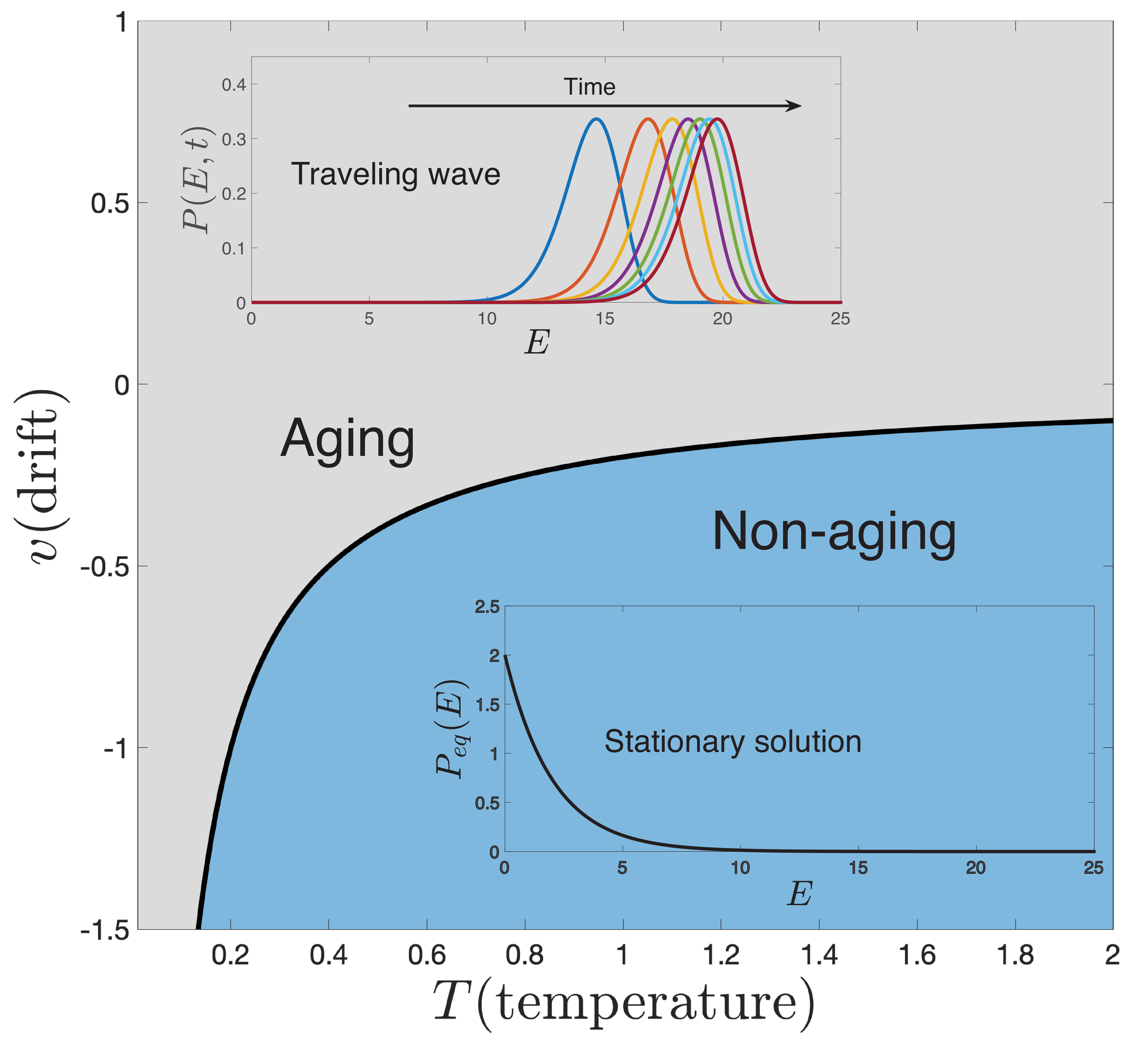}
	\caption{Phase diagram of the aging Maxwell fluid model. In the aging phase in which $v>-D/T$, the yield energy distribution is a traveling wave with a time-dependent traveling speed. Therefore, the system's macroscopic viscosity gradually increases over time. In the non-aging phase, the system exhibits a stationary state without aging.} \label{drift}
\end{figure}

We assume that once a mesoscopic region hops out of its current metastable state, it will reach a new metastable state with a different yield energy. In general, the equation of motion of the yield energy distribution is
\begin{eqnarray}
\frac{\partial P(E,t)}{\partial t} =-P(E,t) e^{-\beta  E}&\nonumber \\
+\int H(E|E^{\prime}) &P(E^{\prime}, t) e^{-\beta E^{\prime}} dE^{\prime}. \label{eq1}
\end{eqnarray}
Here $H(E|E^{\prime})$ is the probability distribution of the yield energy of the new metastable state given the previous state has a yield energy $E^{\prime}$. In the soft glass rheology model \cite{Bouchaud1992, Sollich1997}, $H(E|E^{\prime})$ is assumed to be independent of $E^{\prime}$: the new metastable state is completely independent of the previous metastable state. The soft glassy rheology model predicts a Maxwell fluid state at high temperature and a solid state at low temperature \cite{Sollich1998}, but aging behaviors only exist in the solid phase. Therefore, the soft glassy rheology model is incompatible with experimental observations of increasing viscosity and simultaneously Maxwell fluid viscoelasticity.

Assuming a complete loss of memory of the previous metastable state is a theoretical simplification, making the model analytically solvable. In more realistic models, the correlation between the new metastable state and the previous one should be considered. To incorporate this correlation, we make a simple assumption that the transition kernel $H(E|E^{\prime})=h(E^{\prime}-E)$ and is also short-ranged so that the new metastable states are close to the previous ones in the yield energy space. Therefore, we propose a simple form of Eq. (\ref{eq1}) in the continuum limit as
\begin{equation}
\frac{\partial P(E,t)}{\partial t}=-v \frac{ \partial [P(E, t) e^{-\beta E}]}{\partial E}+D \frac{ \partial^2 [P(E, t) e^{-\beta E}]}{\partial E^2}.\label{diffusion1}
\end{equation}
Here $v=\int h(x)xdx$ is the drift, and $D=\int h(x)x^2dx/2$ is the diffusion constant in the yield energy space. Note that a positive $v$ means a drift towards positive $E$. Eq. (\ref{diffusion1}) resembles the standard diffusion equation, but exhibits a critical difference. For example, even $P(E,t)$ is independent of $E$, the diffusion term has a flux towards the positive direction of $E$ due to the $\exp(-\beta E)$ factor.

Intriguingly, we find that the above equation has an analytical solution that has a traveling wave form $P(E,t)= f(E-v_t t)$ with a time-dependent traveling speed $v_t=T\ln(t)/t$ (Figure \ref{drift}):
\begin{equation}
P(E,t)\sim \frac{e^{(\beta+\frac{v}{D})E}}{t^{1+\frac{v}{D\beta}}}\exp\Big(-\frac{e^{\beta E}}{D\beta^2t}\Big),\label{sol2}
\end{equation}
where we drop the normalization constant, which is straightforward to compute [see derivation details in the Supplementary Materials (SM)]. The traveling wave solutions are confirmed numerically (Figure S1).

The traveling wave solution suggests aging phenomena. We find that the average yield energy increases logarithmically with time $\langle E\rangle = k_B T(\ln(D\beta^2 t))+ \mathrm{const}$ (see the expression of the constant term in SM). Meanwhile, the average relaxation time $\langle \tau_c\rangle = \int e^{\beta E}P(E,t)dE$ increases linearly with time:
\begin{equation}
\langle \tau_c\rangle =\Big(\frac{D}{T^2}+\frac{v}{T}\Big) t \equiv  A t\label{tauc}
\end{equation}
Here we define $A$ as the aging rate of the average relaxation time. The time $t$ here can be considered as the waiting time $t_w$ after the formation of protein condensates. Note that according to Eq. (\ref{tauc}), the average relaxation time $\langle \tau_c(t=0)\rangle=0$, which is a good approximation if $\langle \tau_c (t=0)\rangle$ is negligible compared with the average relaxation time for large $t$. When this approximation is not valid, we find that the yield energy distribution can be well approximated by Eq. (\ref{sol2}) with a constant shift in time (SM).

We find a phase transition from the aging phase to a non-aging phase when $v<-D/T$ (Figure \ref{drift}). In the non-aging phase, the yield energy distribution becomes a stationary solution, $P_{\mathrm{eq}}(E)=-e^{(\beta+v/D)E}/(\beta+v/D)$. Interestingly, we find that the nature of the phase transition appears to be hybrid: the traveling speed changes from zero to a time-dependent value that is independent of the drift; meanwhile, the aging rate of the relaxation time changes continuously, e.g., $A \sim (v-v_c)$ where $v_c=-D/T$. Our model predicts two ways to slow down or even halt aging. One is by increasing the temperature, consistent with experimental observations \cite{Jawerth2020}. The other way is to reduce the drift so that the average yield energy of the new metastable state becomes smaller than the current state.

\begin{figure}[htb!]
	\includegraphics[width=0.4\textwidth]{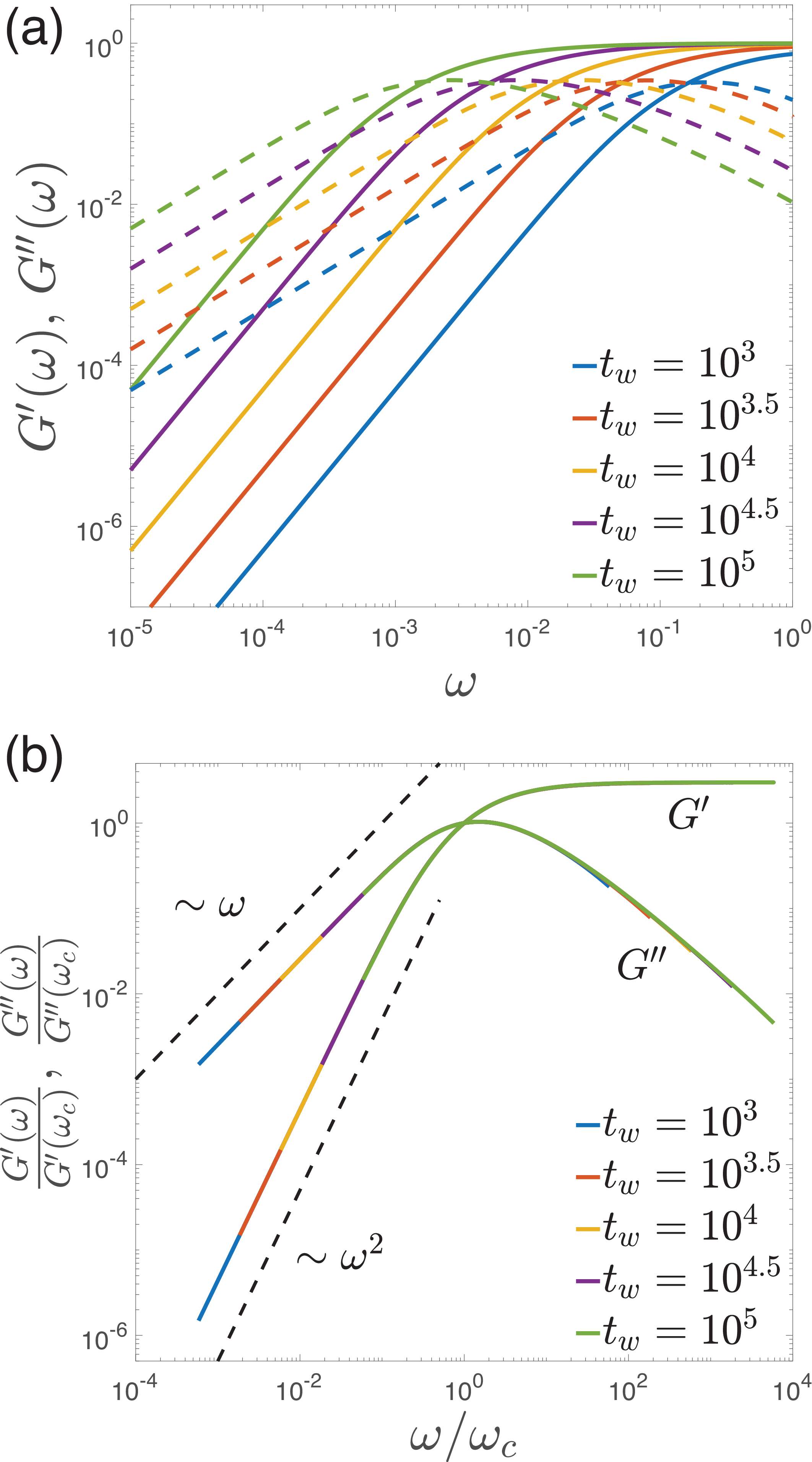}
	\caption{Self-similarity of aging Maxwell fluids at different waiting times.  In this figure, $D=0.2$, $T=2$, $v=0$. (a) The complex modulus as a function of frequency at different waiting times. $G^{\prime}$ is the real part (solid lines), and $G^{\prime\prime}$ is the imaginary part (dashed lines). (b) A rescaled plot of (b) where $\omega_c$ is the characteristic frequency at which $G^{\prime}(\omega_c)=G^{\prime\prime}(\omega_c)$. The same plots with different parameters are shown in Figure S2.}\label{Dv0}
\end{figure}
{\it Self-similarity of complex modulus ---}Imagine we prepare the material at time zero and implement the periodic shear experiment after a waiting time $t_w$. Assuming the yield energy distribution changes little during the experiments (quasi-equilibrium assumption), the complex shear modulus, $G(\omega)=G^{\prime}(\omega)+iG^{\prime\prime}(\omega)$, can be calculated using Eq. (\ref{G}) and Eq. (\ref{sol2}) [Figure \ref{Dv0}(a)]. We rescale the curves by the characteristic frequency $\omega_c$ at which $G^{\prime}(\omega_c)=G^{\prime\prime}(\omega_c)$. The shear modulus curves can be nicely collapsed on a master curve consistent with Maxwell fluid models, with the signature scaling of Maxwell fluids: $G^{\prime}(\omega)\sim \omega^2$ and $G^{\prime\prime}(\omega)\sim \omega$ at low frequency [Figure \ref{Dv0}(b)]. The self-similarity of the complex modulus is robust in the aging phase, as we confirm using multiple sets of parameters, including positive and negative drifts (Figure S2). Furthermore, the characteristic time scale of the whole system, which is defined as $1/\omega_c$, can be well approximated by the average relaxation time $\langle \tau_c\rangle$ (Figure S3). For Maxwell fluid, $\omega_c=k/\eta$ where $k$ is the elastic modulus and $\eta$ is the macroscopic viscosity. Since the elastic constant is constant, our model predicts that the macroscopic viscosity increases linearly with the waiting time between the formation of the protein condensate and the periodic shear experiment.

We now discuss the validity of our quasi-equilibrium assumption. Imagine one does a rheology experiment with time interval $\delta t$ so that the change of $\langle \tau_c\rangle$ in this interval is $\delta \langle \tau_c\rangle =(D\beta^2+v\beta) \delta t$. The condition of $\delta \langle \tau_c\rangle \ll \langle \tau_c\rangle$ becomes $\delta t\ll t_w$. To accurately measure the complex modulus at frequency $\omega$, multiple periods of shear is necessary, $\delta t \gtrsim1/\omega$. Therefore, the minimum frequency we can accurately measure should satisfy $\omega_{\text{min}}\gg 1/t_w$. To measure the complex modulus over a broad range of frequency, we require that $1/t_w \ll 1/\langle \tau_c\rangle$ and according to Eq. (\ref{tauc}), this means $D/T^2 +v/T\ll 1$. We note that experiments have indeed observed that the waiting times are much longer than the characteristic relaxation times ($t_w\gg \langle\tau_c\rangle$) for various protein condensates \cite{Jawerth2020}.

{\it Time-dependent viscosity under constant shear rate ---} Along with the linear viscoelasticity scenario, we also consider the case of constant shear rate $\dot{\gamma}$, which is one of the simplest probes of nonlinear rheology. For idealized Maxwell fluid with a relaxation time $\tau_c$, the stress ($\sigma$)-strain ($\gamma$) curve is universal:
\begin{equation}
\sigma = \dot{\gamma}\tau_c (1-e^{-\frac{\gamma}{\dot{\gamma}\tau_c}}),\label{ideal}
\end{equation}
which means that for two different materials with different relaxation times, their stress-strain curves are identical if they have the same $\dot{\gamma}\tau_c$. However, for aging Maxwell fluids, the average relaxation time depends on the waiting time $t_w$ and the external shear also speeds up the yielding process: the hopping rate of one metastable state with yield energy $E$ now becomes $e^{-\beta (E-\frac{1}{2}l^2)}$ where $l$ is the local strain. For simplicity, we assume that the local strain is reset to $0$ right after the yielding of an element \cite{Sollich1998}. The equation of motion of the distribution of yield energy and local strain is shown in SM [Eq. S(10)].

We present a scaling analysis to derive the long-time behavior of stress-strain response of aging Maxwell fluid. Given a constant shear rate, the typical duration for an element to escape a metastable state with yield energy $E$ becomes $\tau_c(E)\approx\sqrt{2E}/\dot{\gamma}$, namely, the duration after which the elastic energy is equal to the yield energy. Here we neglect the contribution of thermal fluctuation since the average yield energy keeps increasing over time and eventually becomes larger than the thermal energy.

\begin{figure}[htb!]
	\includegraphics[width=0.48\textwidth]{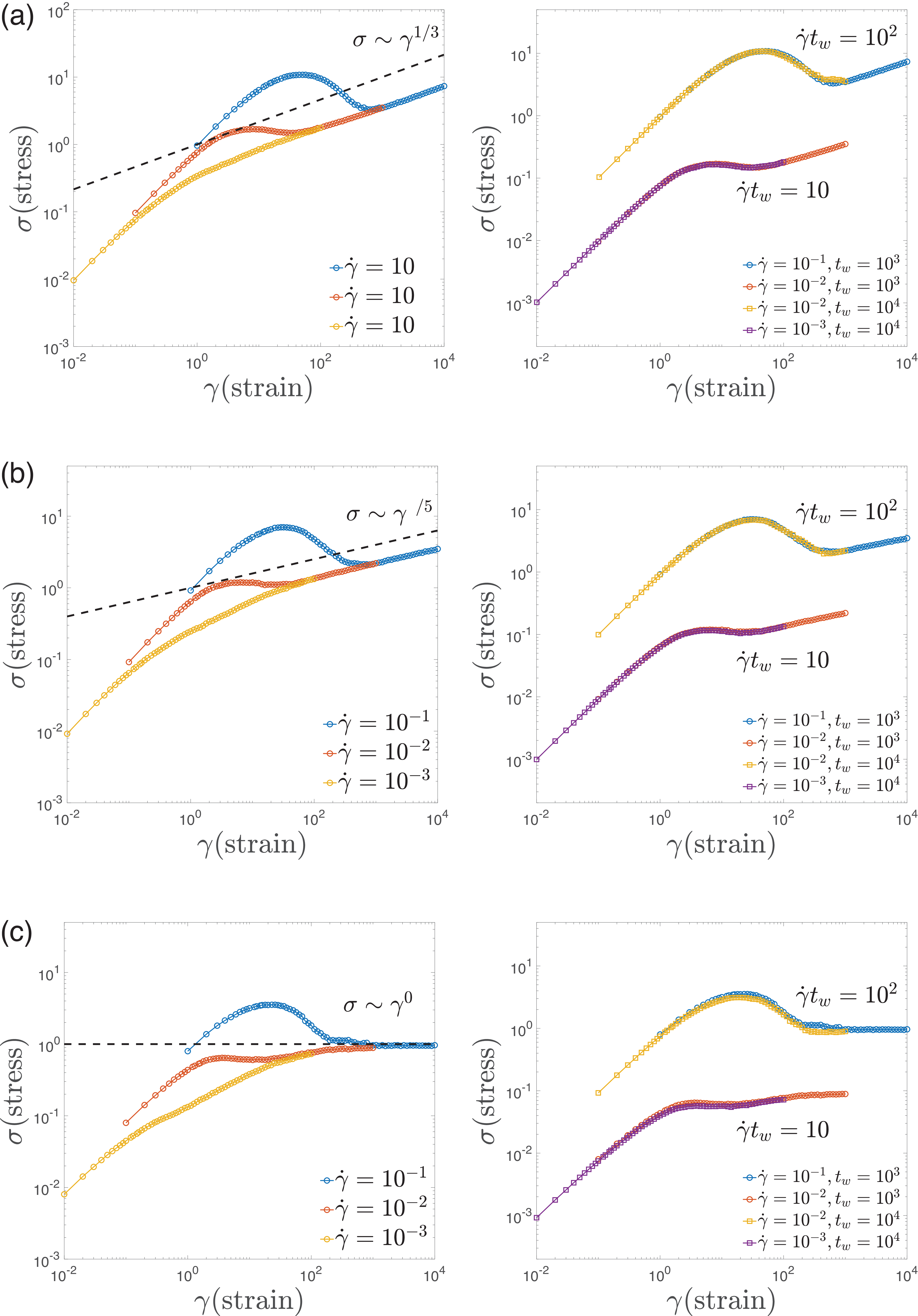}
	\caption{Nonlinear rheology of aging Maxwell fluids under constant shear rate. (a) In the left panel, $D=0.2$, $v=0.1$, $T=1$. The waiting time $t_w=10^3$ with three different shear rates. The dashed line represents the predicted scaling. In the right panel, the four curves have different $\dot{\gamma}$ and $t_w$ as shown in the legend. Stress-strain curves with the same $\dot{\gamma}t_w$ collapse on the same curve. Note that the two curves with $\dot{\gamma}t_w=10$ are rescaled by a factor of $0.1$ in the $y$ axis for a better illustration. (b) Same analysis as (a) with $v=0$ and other parameters the same. (c) Same analysis as (a) with $v=-0.1$ and other parameters the same.}\label{ss}
\end{figure}

We consider that one element has a $1/2$ probability to reach a new metastable state with a yield energy $E+\Delta E_1$ and another $1/2$ probability to reach a new metastable state with a yield energy $E-\Delta E_2$ where $E$ is the yield energy of the current state. We approximate the average yield energy increment due to one hopping out of a metastable state as an average over the two new metastable states weighted by their lifetimes: $\Delta \langle E\rangle =(\Delta E_1 \sqrt{2(E+\Delta E_1)}-\Delta E_2 \sqrt{2(E-\Delta E_2)})/(\sqrt{2(E+\Delta E_1)}+\sqrt{2(E-\Delta E_2)})$, which we expand to second order and obtain $\Delta \langle E\rangle = (\Delta E_1-\Delta E_2)/2 + (\Delta E_1+\Delta E_2)^2/(8E)\approx v+D/E$.
Finally, we obtain the time derivative of the average yield energy as
\begin{equation}
\frac{d\langle E\rangle}{dt} \approx\frac{\Delta \langle E\rangle}{\tau_c(\langle E\rangle )}\approx \dot{\gamma} (\frac{v}{\sqrt{2\langle E\rangle }}+\frac{D}{\sqrt{2}\langle E\rangle^{\frac{3}{2}}}).
\end{equation}
From the above equation and using $\sigma\sim  \tau_c(\langle E\rangle) \dot{\gamma}\sim \sqrt{\langle E\rangle}$, we obtain three different scaling regimes of the stress response:
\begin{eqnarray}
\sigma &\sim (v\dot{\gamma} t_s)^{1/3}, \quad &\mathrm{for} \quad v>0,\nonumber \label{sigma1}\\
\sigma &\sim (D\dot{\gamma} t_s)^{1/5}, \quad &\mathrm{for} \quad v=0,\label{sigma2} \\
\sigma &\sim \sqrt{-D/v}, \quad &\mathrm{for} \quad v<0. \nonumber\label{sigma3}
\end{eqnarray}
Here $t_s$ is the duration of constant shear and the total strain $\gamma = \dot{\gamma} t_s$. Our model predicts that aging Maxwell fluids are rheopectic fluids --- the shear stress increases with time under constant shear rate. We numerically test our predictions by directly simulating a condensate with a large number of elements and compute the shear stress as the average strain over all elements, $\sigma=\langle l\rangle$. The transition kernel $H(E|E^{\prime})=\delta (E-E^{\prime}-\Delta E_1)/2+\delta(E-E^{\prime}+\Delta E_2)/2$ where $\Delta E_1$ and $\Delta E_2$ are chosen such that $v$ and $D$ are equal to the values we set (see simulation details in SM). Our theoretical predictions are nicely confirmed (Figure \ref{ss}). 

Eq. (\ref{sigma2}) shows that the stress $\sigma$ is only a function of the strain $\gamma$ for large $\gamma$, and in the meantime, Eq. (\ref{ideal}) suggest the stress-strain curve of an aging Maxwell fluid should also depend on the product of $\dot{\gamma}$ and its characteristic time scales. Therefore, we propose that the stress-strain curves of aging Maxwell fluids have universal functional forms: $\sigma=F(\gamma, \dot{\gamma}\langle \tau_c\rangle)$ where $\langle \tau_c\rangle$ is the average relaxation time right before the shear, which is approximately constant during the shear experiment. Furthermore, since the average relaxation time is proportional to the waiting time $\langle \tau_c\rangle =At_w$ in our model, we propose a universal shape of the stress-strain curve as $\sigma=F(\gamma,\dot{\gamma}t_w)$. We confirm this functional form numerically and find that the stress-strain curves with equal $\dot{\gamma}t_w$ indeed collapse on the same curve (Figure \ref{ss}).

{\it Discussion ---} In this manuscript, we propose a novel aging Maxwell fluid model in which a protein condensate is subdivided into an ensemble of mesoscopic elements. Each element has a local yield energy which changes after the element escapes the current state and reaches a new metastable state. The critical assumption of the model is to assume that the transition kernel in the yield energy space is short-ranged so that the equation of motion of the yield energy distribution becomes continuous with a drift term and a diffusion term. Surprisingly, the equation has analytical solutions which exhibit phase transition from aging phase to non-aging phase. The order parameter is defined as the aging rate, which changes from zero to a finite value continuously as the drift increases or the temperature decreases. 

In the linear viscoelasticity regime, our model predicts that the complex modulus is self-similar at all times after the formation of condensates. The material behaves as a  Maxwell fluid with a characteristic relaxation time increasing linearly over time. We note that the experiments by Jawerth et al. \cite{Jawerth2020} found that the characteristic relaxation time has a nonlinear scaling relation with the waiting time, which depends on the temperature and salt concentrations. We propose that our model can be generalized to reproduce this nonlinear scaling in multiple possible ways. One possible way is to replace the temperature with an effective temperature that depends on the activity -- the average hopping rate of the whole system. Another possible way is to consider a yield energy-dependent drift [$v(E)$] and/or diffusion constant [$D(E)$]. We expect future works in this direction.

In vivo experiments found that aging processes can be significantly slowed down inside living cells \cite{Shin2017}. While the microscopic mechanisms of how biological activities affect the transition between metastable states are far from clear, our results suggest the effects of biological activities may be increasing the temperature to a higher effective value or biasing the drift towards smaller yield energy.

We also discuss nonlinear rheology with a constant shear rate and find that protein condensates exhibit non-Newtonian rheological behavior with a viscosity increasing over time. This type of complex fluid, called rheopectic fluid, is relatively rare compared with thixotropic fluid, whose viscosity decreases over time under shear. We remark that our theoretical predictions can be experimentally tested. If confirmed, our model suggests that protein condensates exhibit an uncommon rheological property, which may have special biological meaning and practical applications.

Finally, we note that fibrous structures are only occasionally observed on the surface of condensates in the experiments by Jawerth et al. \cite{Jawerth2020}, which appears to be in contrast with other experiments \cite{Murakami2015, Molliex2015, Shin2017}. We propose that the difference could be due to the degree of supersaturation in different experiments. A deep supersaturation can generate solid-like gels with irregular shapes and promote the formation of fibrous structures over time \cite{Shin2017}, which may not be probed in the experiments by Jawerth et al. \cite{Jawerth2020}. Seeking a unified phase diagram of the solid and liquid phases and their corresponding aging behaviors of protein condensates will be an exciting future direction.

We thank Jingxiang Shen, Qiwei Yu, and Lingyu Meng for useful discussions related to this work. J.L. thank the support from the Center for Life Sciences at Peking University.

\bibliography{Lin_Apr_2021}

\begin{thebibliography}{28}%
\makeatletter
\providecommand \@ifxundefined [1]{%
 \@ifx{#1\undefined}
}%
\providecommand \@ifnum [1]{%
 \ifnum #1\expandafter \@firstoftwo
 \else \expandafter \@secondoftwo
 \fi
}%
\providecommand \@ifx [1]{%
 \ifx #1\expandafter \@firstoftwo
 \else \expandafter \@secondoftwo
 \fi
}%
\providecommand \natexlab [1]{#1}%
\providecommand \enquote  [1]{``#1''}%
\providecommand \bibnamefont  [1]{#1}%
\providecommand \bibfnamefont [1]{#1}%
\providecommand \citenamefont [1]{#1}%
\providecommand \href@noop [0]{\@secondoftwo}%
\providecommand \href [0]{\begingroup \@sanitize@url \@href}%
\providecommand \@href[1]{\@@startlink{#1}\@@href}%
\providecommand \@@href[1]{\endgroup#1\@@endlink}%
\providecommand \@sanitize@url [0]{\catcode `\\12\catcode `\$12\catcode
  `\&12\catcode `\#12\catcode `\^12\catcode `\_12\catcode `\%12\relax}%
\providecommand \@@startlink[1]{}%
\providecommand \@@endlink[0]{}%
\providecommand \url  [0]{\begingroup\@sanitize@url \@url }%
\providecommand \@url [1]{\endgroup\@href {#1}{\urlprefix }}%
\providecommand \urlprefix  [0]{URL }%
\providecommand \Eprint [0]{\href }%
\providecommand \doibase [0]{http://dx.doi.org/}%
\providecommand \selectlanguage [0]{\@gobble}%
\providecommand \bibinfo  [0]{\@secondoftwo}%
\providecommand \bibfield  [0]{\@secondoftwo}%
\providecommand \translation [1]{[#1]}%
\providecommand \BibitemOpen [0]{}%
\providecommand \bibitemStop [0]{}%
\providecommand \bibitemNoStop [0]{.\EOS\space}%
\providecommand \EOS [0]{\spacefactor3000\relax}%
\providecommand \BibitemShut  [1]{\csname bibitem#1\endcsname}%
\let\auto@bib@innerbib\@empty
\bibitem [{\citenamefont {Hutchinson}(1995)}]{Hutchinson1995}%
  \BibitemOpen
  \bibfield  {author} {\bibinfo {author} {\bibfnamefont {John~M}\ \bibnamefont
  {Hutchinson}},\ }\bibfield  {title} {\enquote {\bibinfo {title} {Physical
  aging of polymers},}\ }\href@noop {} {\bibfield  {journal} {\bibinfo
  {journal} {Progress in polymer science}\ }\textbf {\bibinfo {volume} {20}},\
  \bibinfo {pages} {703--760} (\bibinfo {year} {1995})}\BibitemShut {NoStop}%
\bibitem [{\citenamefont {Cipelletti}\ \emph {et~al.}(2003)\citenamefont
  {Cipelletti}, \citenamefont {Ramos}, \citenamefont {Manley}, \citenamefont
  {Pitard}, \citenamefont {Weitz}, \citenamefont {Pashkovski},\ and\
  \citenamefont {Johansson}}]{Cipelletti2003}%
  \BibitemOpen
  \bibfield  {author} {\bibinfo {author} {\bibfnamefont {Luca}\ \bibnamefont
  {Cipelletti}}, \bibinfo {author} {\bibfnamefont {Laurence}\ \bibnamefont
  {Ramos}}, \bibinfo {author} {\bibfnamefont {S.}~\bibnamefont {Manley}},
  \bibinfo {author} {\bibfnamefont {E.}~\bibnamefont {Pitard}}, \bibinfo
  {author} {\bibfnamefont {D.~A.}\ \bibnamefont {Weitz}}, \bibinfo {author}
  {\bibfnamefont {Eugene~E.}\ \bibnamefont {Pashkovski}}, \ and\ \bibinfo
  {author} {\bibfnamefont {Marie}\ \bibnamefont {Johansson}},\ }\bibfield
  {title} {\enquote {\bibinfo {title} {Universal non-diffusive slow dynamics in
  aging soft matter},}\ }\href {\doibase 10.1039/B204495A} {\bibfield
  {journal} {\bibinfo  {journal} {Faraday Discuss.}\ }\textbf {\bibinfo
  {volume} {123}},\ \bibinfo {pages} {237--251} (\bibinfo {year}
  {2003})}\BibitemShut {NoStop}%
\bibitem [{\citenamefont {Bonn}\ \emph {et~al.}(2017)\citenamefont {Bonn},
  \citenamefont {Denn}, \citenamefont {Berthier}, \citenamefont {Divoux},\ and\
  \citenamefont {Manneville}}]{Bonn2017}%
  \BibitemOpen
  \bibfield  {author} {\bibinfo {author} {\bibfnamefont {Daniel}\ \bibnamefont
  {Bonn}}, \bibinfo {author} {\bibfnamefont {Morton~M.}\ \bibnamefont {Denn}},
  \bibinfo {author} {\bibfnamefont {Ludovic}\ \bibnamefont {Berthier}},
  \bibinfo {author} {\bibfnamefont {Thibaut}\ \bibnamefont {Divoux}}, \ and\
  \bibinfo {author} {\bibfnamefont {S\'ebastien}\ \bibnamefont {Manneville}},\
  }\bibfield  {title} {\enquote {\bibinfo {title} {Yield stress materials in
  soft condensed matter},}\ }\href {\doibase 10.1103/RevModPhys.89.035005}
  {\bibfield  {journal} {\bibinfo  {journal} {Rev. Mod. Phys.}\ }\textbf
  {\bibinfo {volume} {89}},\ \bibinfo {pages} {035005} (\bibinfo {year}
  {2017})}\BibitemShut {NoStop}%
\bibitem [{\citenamefont {Lahini}\ \emph {et~al.}(2017)\citenamefont {Lahini},
  \citenamefont {Gottesman}, \citenamefont {Amir},\ and\ \citenamefont
  {Rubinstein}}]{Lahini2017}%
  \BibitemOpen
  \bibfield  {author} {\bibinfo {author} {\bibfnamefont {Yoav}\ \bibnamefont
  {Lahini}}, \bibinfo {author} {\bibfnamefont {Omer}\ \bibnamefont
  {Gottesman}}, \bibinfo {author} {\bibfnamefont {Ariel}\ \bibnamefont {Amir}},
  \ and\ \bibinfo {author} {\bibfnamefont {Shmuel~M.}\ \bibnamefont
  {Rubinstein}},\ }\bibfield  {title} {\enquote {\bibinfo {title} {Nonmonotonic
  aging and memory retention in disordered mechanical systems},}\ }\href
  {\doibase 10.1103/PhysRevLett.118.085501} {\bibfield  {journal} {\bibinfo
  {journal} {Phys. Rev. Lett.}\ }\textbf {\bibinfo {volume} {118}},\ \bibinfo
  {pages} {085501} (\bibinfo {year} {2017})}\BibitemShut {NoStop}%
\bibitem [{\citenamefont {Brangwynne}\ \emph {et~al.}(2009)\citenamefont
  {Brangwynne}, \citenamefont {Eckmann}, \citenamefont {Courson}, \citenamefont
  {Rybarska}, \citenamefont {Hoege}, \citenamefont {Gharakhani}, \citenamefont
  {J{\"u}licher},\ and\ \citenamefont {Hyman}}]{Brangwynne2009}%
  \BibitemOpen
  \bibfield  {author} {\bibinfo {author} {\bibfnamefont {Clifford~P.}\
  \bibnamefont {Brangwynne}}, \bibinfo {author} {\bibfnamefont {Christian~R.}\
  \bibnamefont {Eckmann}}, \bibinfo {author} {\bibfnamefont {David~S.}\
  \bibnamefont {Courson}}, \bibinfo {author} {\bibfnamefont {Agata}\
  \bibnamefont {Rybarska}}, \bibinfo {author} {\bibfnamefont {Carsten}\
  \bibnamefont {Hoege}}, \bibinfo {author} {\bibfnamefont {J{\"o}bin}\
  \bibnamefont {Gharakhani}}, \bibinfo {author} {\bibfnamefont {Frank}\
  \bibnamefont {J{\"u}licher}}, \ and\ \bibinfo {author} {\bibfnamefont
  {Anthony~A.}\ \bibnamefont {Hyman}},\ }\bibfield  {title} {\enquote {\bibinfo
  {title} {Germline p granules are liquid droplets that localize by controlled
  dissolution/condensation},}\ }\href@noop {} {\bibfield  {journal} {\bibinfo
  {journal} {Science}\ }\textbf {\bibinfo {volume} {324}},\ \bibinfo {pages}
  {1729--1732} (\bibinfo {year} {2009})}\BibitemShut {NoStop}%
\bibitem [{\citenamefont {Brangwynne}\ \emph {et~al.}(2015)\citenamefont
  {Brangwynne}, \citenamefont {Tompa},\ and\ \citenamefont
  {Pappu}}]{Brangwynne2015}%
  \BibitemOpen
  \bibfield  {author} {\bibinfo {author} {\bibfnamefont {Clifford~P}\
  \bibnamefont {Brangwynne}}, \bibinfo {author} {\bibfnamefont {Peter}\
  \bibnamefont {Tompa}}, \ and\ \bibinfo {author} {\bibfnamefont {Rohit~V}\
  \bibnamefont {Pappu}},\ }\bibfield  {title} {\enquote {\bibinfo {title}
  {Polymer physics of intracellular phase transitions},}\ }\href@noop {}
  {\bibfield  {journal} {\bibinfo  {journal} {Nature Physics}\ }\textbf
  {\bibinfo {volume} {11}},\ \bibinfo {pages} {899--904} (\bibinfo {year}
  {2015})}\BibitemShut {NoStop}%
\bibitem [{\citenamefont {Mao}\ \emph {et~al.}(2019)\citenamefont {Mao},
  \citenamefont {Kuldinow}, \citenamefont {Haataja},\ and\ \citenamefont
  {Ko{\v{s}}mrlj}}]{Mao2019}%
  \BibitemOpen
  \bibfield  {author} {\bibinfo {author} {\bibfnamefont {Sheng}\ \bibnamefont
  {Mao}}, \bibinfo {author} {\bibfnamefont {Derek}\ \bibnamefont {Kuldinow}},
  \bibinfo {author} {\bibfnamefont {Mikko~P}\ \bibnamefont {Haataja}}, \ and\
  \bibinfo {author} {\bibfnamefont {Andrej}\ \bibnamefont {Ko{\v{s}}mrlj}},\
  }\bibfield  {title} {\enquote {\bibinfo {title} {Phase behavior and
  morphology of multicomponent liquid mixtures},}\ }\href@noop {} {\bibfield
  {journal} {\bibinfo  {journal} {Soft Matter}\ }\textbf {\bibinfo {volume}
  {15}},\ \bibinfo {pages} {1297--1311} (\bibinfo {year} {2019})}\BibitemShut
  {NoStop}%
\bibitem [{\citenamefont {Patel}\ \emph {et~al.}(2015)\citenamefont {Patel},
  \citenamefont {Lee}, \citenamefont {Jawerth}, \citenamefont {Maharana},
  \citenamefont {Jahnel}, \citenamefont {Hein}, \citenamefont {Stoynov},
  \citenamefont {Mahamid}, \citenamefont {Saha}, \citenamefont {Franzmann}
  \emph {et~al.}}]{Patel2015}%
  \BibitemOpen
  \bibfield  {author} {\bibinfo {author} {\bibfnamefont {Avinash}\ \bibnamefont
  {Patel}}, \bibinfo {author} {\bibfnamefont {Hyun~O}\ \bibnamefont {Lee}},
  \bibinfo {author} {\bibfnamefont {Louise}\ \bibnamefont {Jawerth}}, \bibinfo
  {author} {\bibfnamefont {Shovamayee}\ \bibnamefont {Maharana}}, \bibinfo
  {author} {\bibfnamefont {Marcus}\ \bibnamefont {Jahnel}}, \bibinfo {author}
  {\bibfnamefont {Marco~Y}\ \bibnamefont {Hein}}, \bibinfo {author}
  {\bibfnamefont {Stoyno}\ \bibnamefont {Stoynov}}, \bibinfo {author}
  {\bibfnamefont {Julia}\ \bibnamefont {Mahamid}}, \bibinfo {author}
  {\bibfnamefont {Shambaditya}\ \bibnamefont {Saha}}, \bibinfo {author}
  {\bibfnamefont {Titus~M}\ \bibnamefont {Franzmann}},  \emph {et~al.},\
  }\bibfield  {title} {\enquote {\bibinfo {title} {A liquid-to-solid phase
  transition of the als protein fus accelerated by disease mutation},}\
  }\href@noop {} {\bibfield  {journal} {\bibinfo  {journal} {Cell}\ }\textbf
  {\bibinfo {volume} {162}},\ \bibinfo {pages} {1066--1077} (\bibinfo {year}
  {2015})}\BibitemShut {NoStop}%
\bibitem [{\citenamefont {Lin}\ \emph {et~al.}(2015)\citenamefont {Lin},
  \citenamefont {Protter}, \citenamefont {Rosen},\ and\ \citenamefont
  {Parker}}]{Lin2015}%
  \BibitemOpen
  \bibfield  {author} {\bibinfo {author} {\bibfnamefont {Yuan}\ \bibnamefont
  {Lin}}, \bibinfo {author} {\bibfnamefont {David~SW}\ \bibnamefont {Protter}},
  \bibinfo {author} {\bibfnamefont {Michael~K}\ \bibnamefont {Rosen}}, \ and\
  \bibinfo {author} {\bibfnamefont {Roy}\ \bibnamefont {Parker}},\ }\bibfield
  {title} {\enquote {\bibinfo {title} {Formation and maturation of
  phase-separated liquid droplets by rna-binding proteins},}\ }\href@noop {}
  {\bibfield  {journal} {\bibinfo  {journal} {Molecular cell}\ }\textbf
  {\bibinfo {volume} {60}},\ \bibinfo {pages} {208--219} (\bibinfo {year}
  {2015})}\BibitemShut {NoStop}%
\bibitem [{\citenamefont {Woodruff}\ \emph {et~al.}(2017)\citenamefont
  {Woodruff}, \citenamefont {Gomes}, \citenamefont {Widlund}, \citenamefont
  {Mahamid}, \citenamefont {Honigmann},\ and\ \citenamefont
  {Hyman}}]{Woodruff2017}%
  \BibitemOpen
  \bibfield  {author} {\bibinfo {author} {\bibfnamefont {Jeffrey~B}\
  \bibnamefont {Woodruff}}, \bibinfo {author} {\bibfnamefont
  {Beatriz~Ferreira}\ \bibnamefont {Gomes}}, \bibinfo {author} {\bibfnamefont
  {Per~O}\ \bibnamefont {Widlund}}, \bibinfo {author} {\bibfnamefont {Julia}\
  \bibnamefont {Mahamid}}, \bibinfo {author} {\bibfnamefont {Alf}\ \bibnamefont
  {Honigmann}}, \ and\ \bibinfo {author} {\bibfnamefont {Anthony~A}\
  \bibnamefont {Hyman}},\ }\bibfield  {title} {\enquote {\bibinfo {title} {The
  centrosome is a selective condensate that nucleates microtubules by
  concentrating tubulin},}\ }\href@noop {} {\bibfield  {journal} {\bibinfo
  {journal} {Cell}\ }\textbf {\bibinfo {volume} {169}},\ \bibinfo {pages}
  {1066--1077} (\bibinfo {year} {2017})}\BibitemShut {NoStop}%
\bibitem [{\citenamefont {Shin}\ and\ \citenamefont
  {Brangwynne}(2017)}]{Shin2017}%
  \BibitemOpen
  \bibfield  {author} {\bibinfo {author} {\bibfnamefont {Yongdae}\ \bibnamefont
  {Shin}}\ and\ \bibinfo {author} {\bibfnamefont {Clifford~P.}\ \bibnamefont
  {Brangwynne}},\ }\bibfield  {title} {\enquote {\bibinfo {title} {Liquid phase
  condensation in cell physiology and disease},}\ }\href@noop {} {\bibfield
  {journal} {\bibinfo  {journal} {Science}\ }\textbf {\bibinfo {volume} {357}}
  (\bibinfo {year} {2017})}\BibitemShut {NoStop}%
\bibitem [{\citenamefont {Banani}\ \emph {et~al.}(2017)\citenamefont {Banani},
  \citenamefont {Lee}, \citenamefont {Hyman},\ and\ \citenamefont
  {Rosen}}]{Banani2017}%
  \BibitemOpen
  \bibfield  {author} {\bibinfo {author} {\bibfnamefont {Salman~F}\
  \bibnamefont {Banani}}, \bibinfo {author} {\bibfnamefont {Hyun~O}\
  \bibnamefont {Lee}}, \bibinfo {author} {\bibfnamefont {Anthony~A}\
  \bibnamefont {Hyman}}, \ and\ \bibinfo {author} {\bibfnamefont {Michael~K}\
  \bibnamefont {Rosen}},\ }\bibfield  {title} {\enquote {\bibinfo {title}
  {Biomolecular condensates: organizers of cellular biochemistry},}\
  }\href@noop {} {\bibfield  {journal} {\bibinfo  {journal} {Nature reviews
  Molecular cell biology}\ }\textbf {\bibinfo {volume} {18}},\ \bibinfo {pages}
  {285--298} (\bibinfo {year} {2017})}\BibitemShut {NoStop}%
\bibitem [{\citenamefont {Franzmann}\ \emph {et~al.}(2018)\citenamefont
  {Franzmann}, \citenamefont {Jahnel}, \citenamefont {Pozniakovsky},
  \citenamefont {Mahamid}, \citenamefont {Holehouse}, \citenamefont
  {N{\"u}ske}, \citenamefont {Richter}, \citenamefont {Baumeister},
  \citenamefont {Grill}, \citenamefont {Pappu}, \citenamefont {Hyman},\ and\
  \citenamefont {Alberti}}]{Franzmann2018}%
  \BibitemOpen
  \bibfield  {author} {\bibinfo {author} {\bibfnamefont {Titus~M.}\
  \bibnamefont {Franzmann}}, \bibinfo {author} {\bibfnamefont {Marcus}\
  \bibnamefont {Jahnel}}, \bibinfo {author} {\bibfnamefont {Andrei}\
  \bibnamefont {Pozniakovsky}}, \bibinfo {author} {\bibfnamefont {Julia}\
  \bibnamefont {Mahamid}}, \bibinfo {author} {\bibfnamefont {Alex~S.}\
  \bibnamefont {Holehouse}}, \bibinfo {author} {\bibfnamefont {Elisabeth}\
  \bibnamefont {N{\"u}ske}}, \bibinfo {author} {\bibfnamefont {Doris}\
  \bibnamefont {Richter}}, \bibinfo {author} {\bibfnamefont {Wolfgang}\
  \bibnamefont {Baumeister}}, \bibinfo {author} {\bibfnamefont {Stephan~W.}\
  \bibnamefont {Grill}}, \bibinfo {author} {\bibfnamefont {Rohit~V.}\
  \bibnamefont {Pappu}}, \bibinfo {author} {\bibfnamefont {Anthony~A.}\
  \bibnamefont {Hyman}}, \ and\ \bibinfo {author} {\bibfnamefont {Simon}\
  \bibnamefont {Alberti}},\ }\bibfield  {title} {\enquote {\bibinfo {title}
  {Phase separation of a yeast prion protein promotes cellular fitness},}\
  }\href@noop {} {\bibfield  {journal} {\bibinfo  {journal} {Science}\ }\textbf
  {\bibinfo {volume} {359}} (\bibinfo {year} {2018})}\BibitemShut {NoStop}%
\bibitem [{\citenamefont {Wang}\ \emph {et~al.}(2018)\citenamefont {Wang},
  \citenamefont {Choi}, \citenamefont {Holehouse}, \citenamefont {Lee},
  \citenamefont {Zhang}, \citenamefont {Jahnel}, \citenamefont {Maharana},
  \citenamefont {Lemaitre}, \citenamefont {Pozniakovsky}, \citenamefont
  {Drechsel} \emph {et~al.}}]{Wang2018}%
  \BibitemOpen
  \bibfield  {author} {\bibinfo {author} {\bibfnamefont {Jie}\ \bibnamefont
  {Wang}}, \bibinfo {author} {\bibfnamefont {Jeong-Mo}\ \bibnamefont {Choi}},
  \bibinfo {author} {\bibfnamefont {Alex~S}\ \bibnamefont {Holehouse}},
  \bibinfo {author} {\bibfnamefont {Hyun~O}\ \bibnamefont {Lee}}, \bibinfo
  {author} {\bibfnamefont {Xiaojie}\ \bibnamefont {Zhang}}, \bibinfo {author}
  {\bibfnamefont {Marcus}\ \bibnamefont {Jahnel}}, \bibinfo {author}
  {\bibfnamefont {Shovamayee}\ \bibnamefont {Maharana}}, \bibinfo {author}
  {\bibfnamefont {R{\'e}gis}\ \bibnamefont {Lemaitre}}, \bibinfo {author}
  {\bibfnamefont {Andrei}\ \bibnamefont {Pozniakovsky}}, \bibinfo {author}
  {\bibfnamefont {David}\ \bibnamefont {Drechsel}},  \emph {et~al.},\
  }\bibfield  {title} {\enquote {\bibinfo {title} {A molecular grammar
  governing the driving forces for phase separation of prion-like rna binding
  proteins},}\ }\href@noop {} {\bibfield  {journal} {\bibinfo  {journal}
  {Cell}\ }\textbf {\bibinfo {volume} {174}},\ \bibinfo {pages} {688--699}
  (\bibinfo {year} {2018})}\BibitemShut {NoStop}%
\bibitem [{\citenamefont {Berry}\ \emph {et~al.}(2018)\citenamefont {Berry},
  \citenamefont {Brangwynne},\ and\ \citenamefont {Haataja}}]{Berry2018}%
  \BibitemOpen
  \bibfield  {author} {\bibinfo {author} {\bibfnamefont {Joel}\ \bibnamefont
  {Berry}}, \bibinfo {author} {\bibfnamefont {Clifford~P}\ \bibnamefont
  {Brangwynne}}, \ and\ \bibinfo {author} {\bibfnamefont {Mikko}\ \bibnamefont
  {Haataja}},\ }\bibfield  {title} {\enquote {\bibinfo {title} {Physical
  principles of intracellular organization via active and passive phase
  transitions},}\ }\href {\doibase 10.1088/1361-6633/aaa61e} {\bibfield
  {journal} {\bibinfo  {journal} {Reports on Progress in Physics}\ }\textbf
  {\bibinfo {volume} {81}},\ \bibinfo {pages} {046601} (\bibinfo {year}
  {2018})}\BibitemShut {NoStop}%
\bibitem [{\citenamefont {Jawerth}\ \emph {et~al.}(2020)\citenamefont
  {Jawerth}, \citenamefont {Fischer-Friedrich}, \citenamefont {Saha},
  \citenamefont {Wang}, \citenamefont {Franzmann}, \citenamefont {Zhang},
  \citenamefont {Sachweh}, \citenamefont {Ruer}, \citenamefont {Ijavi},
  \citenamefont {Saha}, \citenamefont {Mahamid}, \citenamefont {Hyman},\ and\
  \citenamefont {J{\"u}licher}}]{Jawerth2020}%
  \BibitemOpen
  \bibfield  {author} {\bibinfo {author} {\bibfnamefont {Louise}\ \bibnamefont
  {Jawerth}}, \bibinfo {author} {\bibfnamefont {Elisabeth}\ \bibnamefont
  {Fischer-Friedrich}}, \bibinfo {author} {\bibfnamefont {Suropriya}\
  \bibnamefont {Saha}}, \bibinfo {author} {\bibfnamefont {Jie}\ \bibnamefont
  {Wang}}, \bibinfo {author} {\bibfnamefont {Titus}\ \bibnamefont {Franzmann}},
  \bibinfo {author} {\bibfnamefont {Xiaojie}\ \bibnamefont {Zhang}}, \bibinfo
  {author} {\bibfnamefont {Jenny}\ \bibnamefont {Sachweh}}, \bibinfo {author}
  {\bibfnamefont {Martine}\ \bibnamefont {Ruer}}, \bibinfo {author}
  {\bibfnamefont {Mahdiye}\ \bibnamefont {Ijavi}}, \bibinfo {author}
  {\bibfnamefont {Shambaditya}\ \bibnamefont {Saha}}, \bibinfo {author}
  {\bibfnamefont {Julia}\ \bibnamefont {Mahamid}}, \bibinfo {author}
  {\bibfnamefont {Anthony~A.}\ \bibnamefont {Hyman}}, \ and\ \bibinfo {author}
  {\bibfnamefont {Frank}\ \bibnamefont {J{\"u}licher}},\ }\bibfield  {title}
  {\enquote {\bibinfo {title} {Protein condensates as aging maxwell fluids},}\
  }\href {\doibase 10.1126/science.aaw4951} {\bibfield  {journal} {\bibinfo
  {journal} {Science}\ }\textbf {\bibinfo {volume} {370}},\ \bibinfo {pages}
  {1317--1323} (\bibinfo {year} {2020})}\BibitemShut {NoStop}%
\bibitem [{\citenamefont {Jawerth}\ \emph {et~al.}(2018)\citenamefont
  {Jawerth}, \citenamefont {Ijavi}, \citenamefont {Ruer}, \citenamefont {Saha},
  \citenamefont {Jahnel}, \citenamefont {Hyman}, \citenamefont {J\"ulicher},\
  and\ \citenamefont {Fischer-Friedrich}}]{Jawerth2018}%
  \BibitemOpen
  \bibfield  {author} {\bibinfo {author} {\bibfnamefont {Louise~M.}\
  \bibnamefont {Jawerth}}, \bibinfo {author} {\bibfnamefont {Mahdiye}\
  \bibnamefont {Ijavi}}, \bibinfo {author} {\bibfnamefont {Martine}\
  \bibnamefont {Ruer}}, \bibinfo {author} {\bibfnamefont {Shambaditya}\
  \bibnamefont {Saha}}, \bibinfo {author} {\bibfnamefont {Marcus}\ \bibnamefont
  {Jahnel}}, \bibinfo {author} {\bibfnamefont {Anthony~A.}\ \bibnamefont
  {Hyman}}, \bibinfo {author} {\bibfnamefont {Frank}\ \bibnamefont
  {J\"ulicher}}, \ and\ \bibinfo {author} {\bibfnamefont {Elisabeth}\
  \bibnamefont {Fischer-Friedrich}},\ }\bibfield  {title} {\enquote {\bibinfo
  {title} {Salt-dependent rheology and surface tension of protein condensates
  using optical traps},}\ }\href {\doibase 10.1103/PhysRevLett.121.258101}
  {\bibfield  {journal} {\bibinfo  {journal} {Phys. Rev. Lett.}\ }\textbf
  {\bibinfo {volume} {121}},\ \bibinfo {pages} {258101} (\bibinfo {year}
  {2018})}\BibitemShut {NoStop}%
\bibitem [{\citenamefont {Sollich}\ \emph {et~al.}(1997)\citenamefont
  {Sollich}, \citenamefont {Lequeux}, \citenamefont {H\'ebraud},\ and\
  \citenamefont {Cates}}]{Sollich1997}%
  \BibitemOpen
  \bibfield  {author} {\bibinfo {author} {\bibfnamefont {Peter}\ \bibnamefont
  {Sollich}}, \bibinfo {author} {\bibfnamefont {Fran\ifmmode
  \mbox{\c{c}}\else~\c{c}\fi{}ois}\ \bibnamefont {Lequeux}}, \bibinfo {author}
  {\bibfnamefont {Pascal}\ \bibnamefont {H\'ebraud}}, \ and\ \bibinfo {author}
  {\bibfnamefont {Michael~E.}\ \bibnamefont {Cates}},\ }\bibfield  {title}
  {\enquote {\bibinfo {title} {Rheology of soft glassy materials},}\ }\href
  {\doibase 10.1103/PhysRevLett.78.2020} {\bibfield  {journal} {\bibinfo
  {journal} {Phys. Rev. Lett.}\ }\textbf {\bibinfo {volume} {78}},\ \bibinfo
  {pages} {2020--2023} (\bibinfo {year} {1997})}\BibitemShut {NoStop}%
\bibitem [{\citenamefont {Sollich}(1998)}]{Sollich1998}%
  \BibitemOpen
  \bibfield  {author} {\bibinfo {author} {\bibfnamefont {Peter}\ \bibnamefont
  {Sollich}},\ }\bibfield  {title} {\enquote {\bibinfo {title} {Rheological
  constitutive equation for a model of soft glassy materials},}\ }\href
  {\doibase 10.1103/PhysRevE.58.738} {\bibfield  {journal} {\bibinfo  {journal}
  {Phys. Rev. E}\ }\textbf {\bibinfo {volume} {58}},\ \bibinfo {pages}
  {738--759} (\bibinfo {year} {1998})}\BibitemShut {NoStop}%
\bibitem [{\citenamefont {Fielding}\ \emph {et~al.}(2000)\citenamefont
  {Fielding}, \citenamefont {Sollich},\ and\ \citenamefont
  {Cates}}]{Fielding2000}%
  \BibitemOpen
  \bibfield  {author} {\bibinfo {author} {\bibfnamefont {S.~M.}\ \bibnamefont
  {Fielding}}, \bibinfo {author} {\bibfnamefont {P.}~\bibnamefont {Sollich}}, \
  and\ \bibinfo {author} {\bibfnamefont {M.~E.}\ \bibnamefont {Cates}},\
  }\bibfield  {title} {\enquote {\bibinfo {title} {Aging and rheology in soft
  materials},}\ }\href@noop {} {\bibfield  {journal} {\bibinfo  {journal}
  {Journal of Rheology}\ }\textbf {\bibinfo {volume} {44}},\ \bibinfo {pages}
  {323--369} (\bibinfo {year} {2000})}\BibitemShut {NoStop}%
\bibitem [{\citenamefont {Derec}\ \emph {et~al.}(2001)\citenamefont {Derec},
  \citenamefont {Ajdari},\ and\ \citenamefont {Lequeux}}]{Derec2001}%
  \BibitemOpen
  \bibfield  {author} {\bibinfo {author} {\bibfnamefont {Caroline}\
  \bibnamefont {Derec}}, \bibinfo {author} {\bibfnamefont {Armand}\
  \bibnamefont {Ajdari}}, \ and\ \bibinfo {author} {\bibfnamefont {Francois}\
  \bibnamefont {Lequeux}},\ }\bibfield  {title} {\enquote {\bibinfo {title}
  {Rheology and aging: A simple approach},}\ }\href@noop {} {\bibfield
  {journal} {\bibinfo  {journal} {The European Physical Journal E}\ }\textbf
  {\bibinfo {volume} {4}},\ \bibinfo {pages} {355--361} (\bibinfo {year}
  {2001})}\BibitemShut {NoStop}%
\bibitem [{\citenamefont {Fielding}\ \emph {et~al.}(2009)\citenamefont
  {Fielding}, \citenamefont {Cates},\ and\ \citenamefont
  {Sollich}}]{Fielding2009}%
  \BibitemOpen
  \bibfield  {author} {\bibinfo {author} {\bibfnamefont {S.~M.}\ \bibnamefont
  {Fielding}}, \bibinfo {author} {\bibfnamefont {M.~E.}\ \bibnamefont {Cates}},
  \ and\ \bibinfo {author} {\bibfnamefont {P.}~\bibnamefont {Sollich}},\
  }\bibfield  {title} {\enquote {\bibinfo {title} {Shear banding{,} aging and
  noise dynamics in soft glassy materials},}\ }\href {\doibase
  10.1039/B812394M} {\bibfield  {journal} {\bibinfo  {journal} {Soft Matter}\
  }\textbf {\bibinfo {volume} {5}},\ \bibinfo {pages} {2378--2382} (\bibinfo
  {year} {2009})}\BibitemShut {NoStop}%
\bibitem [{\citenamefont {Sollich}\ \emph {et~al.}(2017)\citenamefont
  {Sollich}, \citenamefont {Olivier},\ and\ \citenamefont
  {Bresch}}]{Sollich2017}%
  \BibitemOpen
  \bibfield  {author} {\bibinfo {author} {\bibfnamefont {Peter}\ \bibnamefont
  {Sollich}}, \bibinfo {author} {\bibfnamefont {Julien}\ \bibnamefont
  {Olivier}}, \ and\ \bibinfo {author} {\bibfnamefont {Didier}\ \bibnamefont
  {Bresch}},\ }\bibfield  {title} {\enquote {\bibinfo {title} {Aging and linear
  response in the h{\'{e}}braud{\textendash}lequeux model for amorphous
  rheology},}\ }\href@noop {} {\bibfield  {journal} {\bibinfo  {journal}
  {Journal of Physics A: Mathematical and Theoretical}\ }\textbf {\bibinfo
  {volume} {50}},\ \bibinfo {pages} {165002} (\bibinfo {year}
  {2017})}\BibitemShut {NoStop}%
\bibitem [{\citenamefont {Parley}\ \emph {et~al.}(2020)\citenamefont {Parley},
  \citenamefont {Fielding},\ and\ \citenamefont {Sollich}}]{Parley2020}%
  \BibitemOpen
  \bibfield  {author} {\bibinfo {author} {\bibfnamefont {Jack~T.}\ \bibnamefont
  {Parley}}, \bibinfo {author} {\bibfnamefont {Suzanne~M.}\ \bibnamefont
  {Fielding}}, \ and\ \bibinfo {author} {\bibfnamefont {Peter}\ \bibnamefont
  {Sollich}},\ }\bibfield  {title} {\enquote {\bibinfo {title} {Aging in a mean
  field elastoplastic model of amorphous solids},}\ }\href@noop {} {\bibfield
  {journal} {\bibinfo  {journal} {Physics of Fluids}\ }\textbf {\bibinfo
  {volume} {32}},\ \bibinfo {pages} {127104} (\bibinfo {year}
  {2020})}\BibitemShut {NoStop}%
\bibitem [{\citenamefont {Chebaro}\ \emph {et~al.}(2015)\citenamefont
  {Chebaro}, \citenamefont {Ballard}, \citenamefont {Chakraborty},\ and\
  \citenamefont {Wales}}]{Chebaro2015}%
  \BibitemOpen
  \bibfield  {author} {\bibinfo {author} {\bibfnamefont {Yassmine}\
  \bibnamefont {Chebaro}}, \bibinfo {author} {\bibfnamefont {Andrew~J}\
  \bibnamefont {Ballard}}, \bibinfo {author} {\bibfnamefont {Debayan}\
  \bibnamefont {Chakraborty}}, \ and\ \bibinfo {author} {\bibfnamefont
  {David~J}\ \bibnamefont {Wales}},\ }\bibfield  {title} {\enquote {\bibinfo
  {title} {Intrinsically disordered energy landscapes},}\ }\href@noop {}
  {\bibfield  {journal} {\bibinfo  {journal} {Scientific reports}\ }\textbf
  {\bibinfo {volume} {5}},\ \bibinfo {pages} {1--12} (\bibinfo {year}
  {2015})}\BibitemShut {NoStop}%
\bibitem [{\citenamefont {Bouchaud}(1992)}]{Bouchaud1992}%
  \BibitemOpen
  \bibfield  {author} {\bibinfo {author} {\bibfnamefont {Jean-Philippe}\
  \bibnamefont {Bouchaud}},\ }\bibfield  {title} {\enquote {\bibinfo {title}
  {Weak ergodicity breaking and aging in disordered systems},}\ }\href@noop {}
  {\bibfield  {journal} {\bibinfo  {journal} {Journal de Physique I}\ }\textbf
  {\bibinfo {volume} {2}},\ \bibinfo {pages} {1705--1713} (\bibinfo {year}
  {1992})}\BibitemShut {NoStop}%
\bibitem [{\citenamefont {Murakami}\ \emph {et~al.}(2015)\citenamefont
  {Murakami}, \citenamefont {Qamar}, \citenamefont {Lin}, \citenamefont
  {Schierle}, \citenamefont {Rees}, \citenamefont {Miyashita}, \citenamefont
  {Costa}, \citenamefont {Dodd}, \citenamefont {Chan}, \citenamefont {Michel}
  \emph {et~al.}}]{Murakami2015}%
  \BibitemOpen
  \bibfield  {author} {\bibinfo {author} {\bibfnamefont {Tetsuro}\ \bibnamefont
  {Murakami}}, \bibinfo {author} {\bibfnamefont {Seema}\ \bibnamefont {Qamar}},
  \bibinfo {author} {\bibfnamefont {Julie~Qiaojin}\ \bibnamefont {Lin}},
  \bibinfo {author} {\bibfnamefont {Gabriele S~Kaminski}\ \bibnamefont
  {Schierle}}, \bibinfo {author} {\bibfnamefont {Eric}\ \bibnamefont {Rees}},
  \bibinfo {author} {\bibfnamefont {Akinori}\ \bibnamefont {Miyashita}},
  \bibinfo {author} {\bibfnamefont {Ana~R}\ \bibnamefont {Costa}}, \bibinfo
  {author} {\bibfnamefont {Roger~B}\ \bibnamefont {Dodd}}, \bibinfo {author}
  {\bibfnamefont {Fiona~TS}\ \bibnamefont {Chan}}, \bibinfo {author}
  {\bibfnamefont {Claire~H}\ \bibnamefont {Michel}},  \emph {et~al.},\
  }\bibfield  {title} {\enquote {\bibinfo {title} {Als/ftd mutation-induced
  phase transition of fus liquid droplets and reversible hydrogels into
  irreversible hydrogels impairs rnp granule function},}\ }\href@noop {}
  {\bibfield  {journal} {\bibinfo  {journal} {Neuron}\ }\textbf {\bibinfo
  {volume} {88}},\ \bibinfo {pages} {678--690} (\bibinfo {year}
  {2015})}\BibitemShut {NoStop}%
\bibitem [{\citenamefont {Molliex}\ \emph {et~al.}(2015)\citenamefont
  {Molliex}, \citenamefont {Temirov}, \citenamefont {Lee}, \citenamefont
  {Coughlin}, \citenamefont {Kanagaraj}, \citenamefont {Kim}, \citenamefont
  {Mittag},\ and\ \citenamefont {Taylor}}]{Molliex2015}%
  \BibitemOpen
  \bibfield  {author} {\bibinfo {author} {\bibfnamefont {Amandine}\
  \bibnamefont {Molliex}}, \bibinfo {author} {\bibfnamefont {Jamshid}\
  \bibnamefont {Temirov}}, \bibinfo {author} {\bibfnamefont {Jihun}\
  \bibnamefont {Lee}}, \bibinfo {author} {\bibfnamefont {Maura}\ \bibnamefont
  {Coughlin}}, \bibinfo {author} {\bibfnamefont {Anderson~P}\ \bibnamefont
  {Kanagaraj}}, \bibinfo {author} {\bibfnamefont {Hong~Joo}\ \bibnamefont
  {Kim}}, \bibinfo {author} {\bibfnamefont {Tanja}\ \bibnamefont {Mittag}}, \
  and\ \bibinfo {author} {\bibfnamefont {J~Paul}\ \bibnamefont {Taylor}},\
  }\bibfield  {title} {\enquote {\bibinfo {title} {Phase separation by low
  complexity domains promotes stress granule assembly and drives pathological
  fibrillization},}\ }\href@noop {} {\bibfield  {journal} {\bibinfo  {journal}
  {Cell}\ }\textbf {\bibinfo {volume} {163}},\ \bibinfo {pages} {123--133}
  (\bibinfo {year} {2015})}\BibitemShut {NoStop}%
\end{thebibliography}%


\begin{thebibliography}{0}%
\makeatletter
\providecommand \@ifxundefined [1]{%
 \@ifx{#1\undefined}
}%
\providecommand \@ifnum [1]{%
 \ifnum #1\expandafter \@firstoftwo
 \else \expandafter \@secondoftwo
 \fi
}%
\providecommand \@ifx [1]{%
 \ifx #1\expandafter \@firstoftwo
 \else \expandafter \@secondoftwo
 \fi
}%
\providecommand \natexlab [1]{#1}%
\providecommand \enquote  [1]{``#1''}%
\providecommand \bibnamefont  [1]{#1}%
\providecommand \bibfnamefont [1]{#1}%
\providecommand \citenamefont [1]{#1}%
\providecommand \href@noop [0]{\@secondoftwo}%
\providecommand \href [0]{\begingroup \@sanitize@url \@href}%
\providecommand \@href[1]{\@@startlink{#1}\@@href}%
\providecommand \@@href[1]{\endgroup#1\@@endlink}%
\providecommand \@sanitize@url [0]{\catcode `\\12\catcode `\$12\catcode
  `\&12\catcode `\#12\catcode `\^12\catcode `\_12\catcode `\%12\relax}%
\providecommand \@@startlink[1]{}%
\providecommand \@@endlink[0]{}%
\providecommand \url  [0]{\begingroup\@sanitize@url \@url }%
\providecommand \@url [1]{\endgroup\@href {#1}{\urlprefix }}%
\providecommand \urlprefix  [0]{URL }%
\providecommand \Eprint [0]{\href }%
\providecommand \doibase [0]{http://dx.doi.org/}%
\providecommand \selectlanguage [0]{\@gobble}%
\providecommand \bibinfo  [0]{\@secondoftwo}%
\providecommand \bibfield  [0]{\@secondoftwo}%
\providecommand \translation [1]{[#1]}%
\providecommand \BibitemOpen [0]{}%
\providecommand \bibitemStop [0]{}%
\providecommand \bibitemNoStop [0]{.\EOS\space}%
\providecommand \EOS [0]{\spacefactor3000\relax}%
\providecommand \BibitemShut  [1]{\csname bibitem#1\endcsname}%
\let\auto@bib@innerbib\@empty
\end{thebibliography}%
\end{document}


\title{Supplementary Materials}
\author{Jie Lin}
\affiliation{Center for Quantitative Biology and Peking-Tsinghua Center for Life Sciences, Academy for Advanced Interdisciplinary Studies, Peking University, Beijing 100871, China}
\date{\today}

\maketitle
\subsection{Derivation of the traveling wave solution}

For the equation of motion of the yield energy distribution,
\begin{equation}
\frac{\partial P(E,t)}{\partial t}=-v \frac{ \partial [P(E, t) e^{-\beta E}]}{\partial E}+D \frac{ \partial^2 [P(E, t) e^{-\beta E}]}{\partial E^2},\label{pet}
\end{equation}

we consider an ansatz of the solution:
\begin{equation}
P(E,t)=F(E-T\ln (t))\equiv F(z).
\end{equation}
Therefore, Eq. (\ref{pet}) becomes
\begin{equation}
D\partial_z^2 F(z)-(v+2D\beta-\frac{e^{\beta z}}{\beta})\partial_z F + (v\beta +D\beta^2)F=0.
\end{equation}
We find that the above equation has the following analytical solution:
\begin{equation}
F(z)=\frac{\beta}{(D\beta^2)^{1+\frac{v}{D\beta}}\Gamma(1+\frac{v}{D\beta})} e^{(\beta+\frac{v}{D})z}\exp\big(-\frac{e^{\beta z}}{D\beta^2}\big)\label{solFz}
\end{equation}
where the pre-factor is the normalization constant. Therefore, the solution to Eq. (\ref{pet}) becomes
\begin{equation}
P(E,t)=\frac{\beta}{(D\beta^2)^{1+\frac{v}{D\beta}}\Gamma(1+\frac{v}{D\beta}) t^{1+\frac{v}{D\beta}}} e^{(\beta+\frac{v}{D})E}\exp\big(-\frac{e^{\beta E}}{D\beta^2t}\big)\label{sol1}
\end{equation}

Using Eq. (\ref{sol1}), we find the average yield energy as
\begin{equation}
\langle E\rangle=\frac{1}{\beta \Gamma(1+\frac{v}{D\beta})}\int_{\frac{1}{D\beta^2t}}^{\infty} x^{\frac{v}{D\beta}}e^{-x}\ln(xD\beta^2t)dx =T\Big(\frac{I(\frac{v}{D\beta})}{\Gamma(1+\frac{v}{D\beta})} + \ln(D\beta^2t)\Big).\label{Eave}
\end{equation}
where $I(\frac{v}{D\beta})=\int_0^{\infty} y^{\frac{v}{D\beta}} \ln(y) e^{-y}dy$. Similarly, we find the average relaxation time as
\begin{equation}
\langle \tau_c\rangle = D\beta^2 t\int_{\frac{1}{D\beta^2t}}^{\infty} x^{1+\frac{v}{D\beta}} e^{-x}dx  = (D\beta^2+v\beta)t.\label{tauc}
\end{equation}
Note that in deriving Eqs. (\ref{Eave}, \ref{tauc}), we assume that $D\beta^2 t\gg 1$.

\begin{figure}[htb!]
	\includegraphics[width=0.7\textwidth]{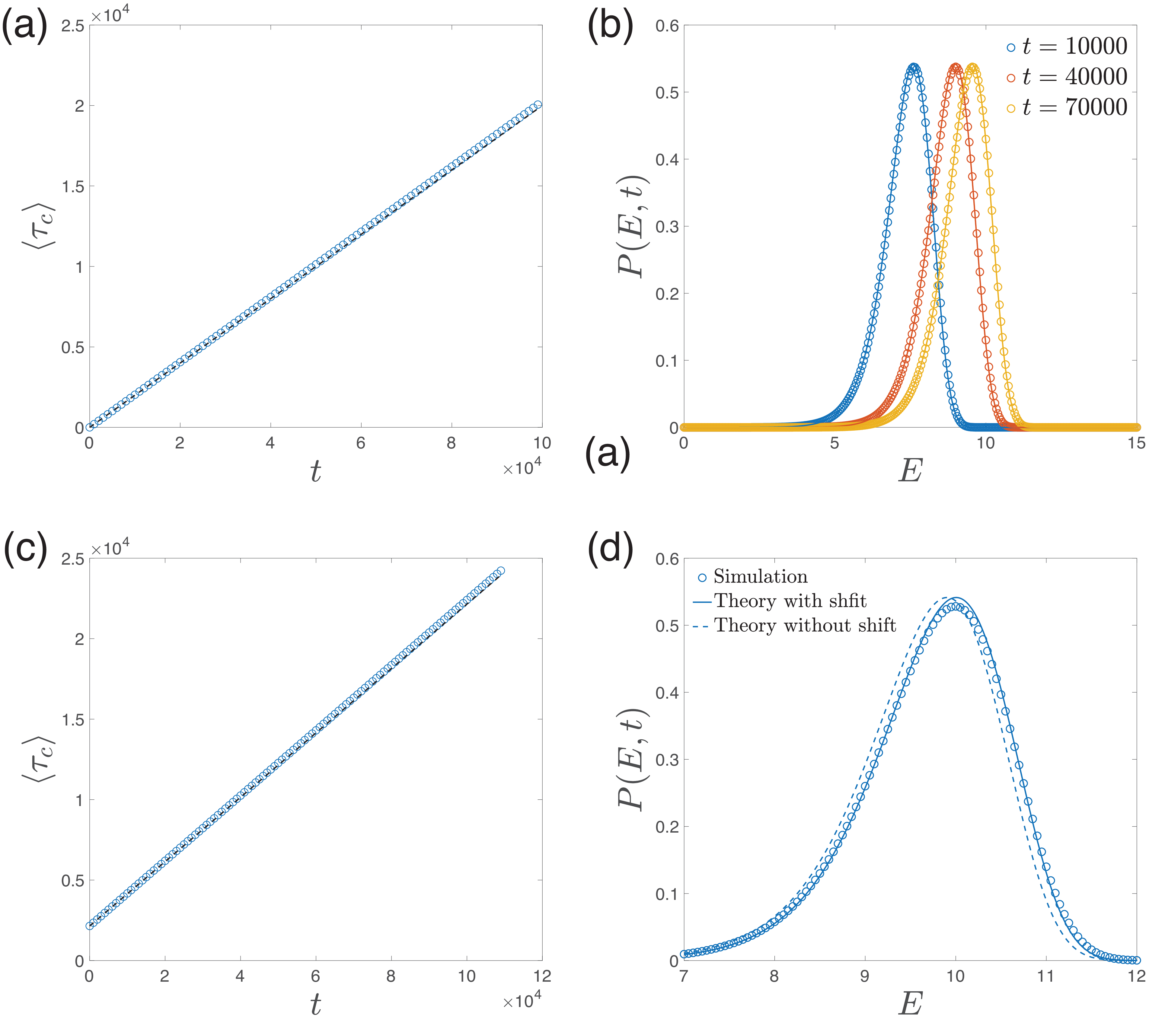}
	\caption{Numerical test of the traveling eave solution. In this figure, $D=0.1$, $v=0.1$, $T=1$. (a) The average relaxation time as a function of time. The circles are simulation results and the dashed line is the theoretical prediction Eq. (\ref{tauc}). (b) The yield energy distribution at different times. The circles are simulation results and the solid lines are the theoretical predictions, Eq. (\ref{sol1}). In (a) and (b), $P(E,t=0)=1$ for $0<E<1$ and $0$ otherwise. (c) The average relaxation time as a function of time. The dashed line is $A(t+t_0)$ where $At_0$ is the average relaxation time at $t=0$. (d) The yield energy distribution at $t=10^5$. The dashed line is Eq. (\ref{sol1}) and the solid line is Eq. (\ref{sol1}) with $t$ replaced by $t+t_0$. In (c) and (d), $P(E,t=0)=1/10$ for $0<E<10$ and $0$ otherwise.}\label{test}
\end{figure}

Alternatively, one can solve Eq. (\ref{pet}) by re-defining the variable $y=e^{\beta E}$ so that $P(y,t)=P(E,t)/(\beta y)$. Therefore, Eq. (\ref{pet}) becomes
\begin{equation}
\partial_t P(y,t)=(D\beta^2-\beta v)\partial_y P(y,t) + D\beta^2 y\partial_y^2 P(y,t).
\end{equation}
We find that the above equation has the following solution:
\begin{equation}
P(y,t)=\frac{1}{(D\beta^2t)^{1+\frac{v}{D\beta}}\Gamma(1+\frac{v}{D\beta})} y^{\frac{v}{D\beta}}e^{-\frac{y}{D\beta^2 t}},
\end{equation}
which leads to the same solution as Eq. (\ref{sol1}).

\subsection{Numerical confirmation of the traveling wave solution}

We numerically solve Eq. (\ref{pet}) by discretizing the yield energy space with $\delta E=0.05$. We update the new yield energy distribution with a time step $\delta t=0.01$ and keep the boundary condition so that $P(E<0, t)=0$. We choose the initial condition such that $P(E,t=0)=1$ for $0<E<1$ and find that the numerical solutions match the theoretical predictions very well [Figure \ref{test}(a,b)].

We also take a wider initial distribution such that $P(E,t=0)=1/10$ for $0<E<10$. In this case $\langle \tau_c \rangle $ at $t=0$ is not negligible and the average relaxation time can be expressed as $\langle \tau_c\rangle =A(t+t_0)$ where $At_0$ is the average relaxation time at $t=0$ [Figure \ref{test}(c)]. We find that $P(E,t)$ can be well approximated by Eq. (\ref{sol1}) with $t$ replaced by $t+t_0$ [Figure \ref{test}(d)].

\subsection{Numerical simulations of constant shear rate}
In the case of nonlinear rheology, the system is described by the yield energy and local strain distribution whose equation of motion is
\begin{equation}
\partial_t P(E,l,t)= -P(E,t) e^{-\beta  (E-\frac{1}{2}l^2)} +\delta(l) \int H(E|E^{\prime}) P(E^{\prime}, l^{\prime}, t) e^{-\beta (E^{\prime}-\frac{1}{2}l^{\prime 2})} dE^{\prime}dl^{\prime}. \label{eq2}
\end{equation}

To test our predictions regarding the rheological response of aging Maxwell fluids under constant shear rate, we simulate a protein condensate with $N$ elements. The yield energy distribution at time $0$, $P(E,t=0)$ is uniformly distributed between $0$ and $E_{m}$ and all elements' local strains are zero. The hopping rate for each element follows $R(E)=\exp(-(E-l^2/2)/T)$ and once an element with yield energy $E$ yields, it will reaches a new metastable state and the transition kernel is $H(E|E^{\prime}) =( \delta (E-E^{\prime}-\Delta E_1) + \delta (E-E^{\prime}+\Delta E_2))/2$ where $\Delta E_1$ and $\Delta E_2$ are constant. For given $v$ and $D$, we choose $\Delta E_1$ and $\Delta E_2$ such that
\begin{eqnarray}
v=\Delta E_1-\Delta E_2, \\
D=\frac{\Delta E_1^2+\Delta E_2^2}{4}.
\end{eqnarray}
In Figure 4 of the main text, we set $E_m=1$, $D=0.2$, $T=1$ and $v=-0.1, 0, 0.1$.

\begin{figure}[htb!]
	\includegraphics[width=0.7\textwidth]{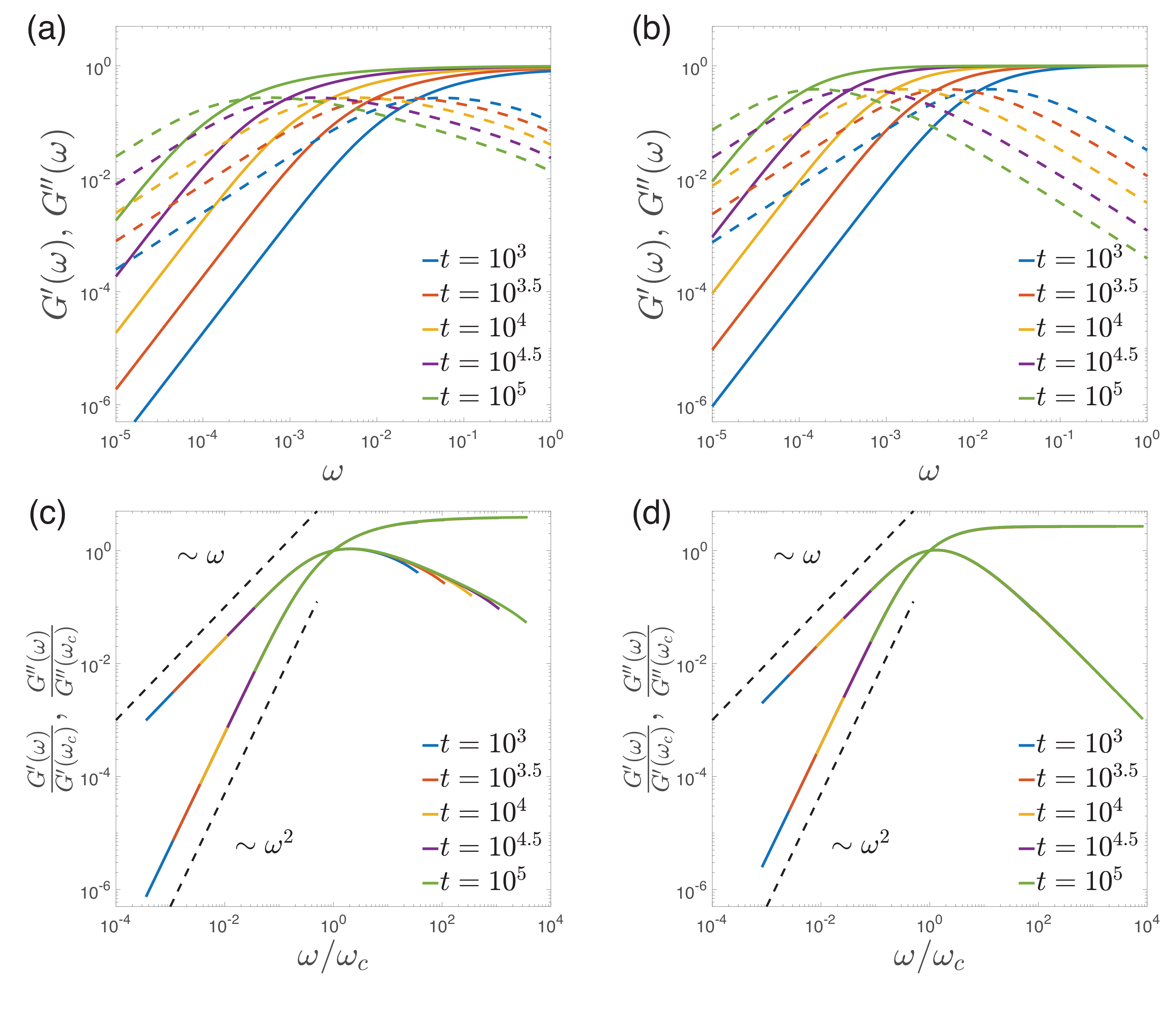}
	\caption{The complex shear modulus for positive and negative drift. (a) The complex moduli as function of frequency at different waiting times. $G^{\prime}$ is the real part (solid lines) and $G^{\prime\prime}$ is the imaginary part (dashed lines). In this panel, $D=0.2$, $T=2$, $v=0.05$. (b) The same analysis as (a) but with $v=-0.05$. (c) A rescaled plot of (a) where $\omega_c$ is the characteristic frequency at which $G^{\prime}(\omega_c)=G^{\prime\prime}(\omega_c)$. (d) A rescaled plot of (b) where $\omega_c$ is the characteristic frequency at which $G^{\prime}(\omega_c)=G^{\prime\prime}(\omega_c)$. }\label{collapse}
\end{figure}

\begin{figure}[htb!]
	\includegraphics[width=0.45\textwidth]{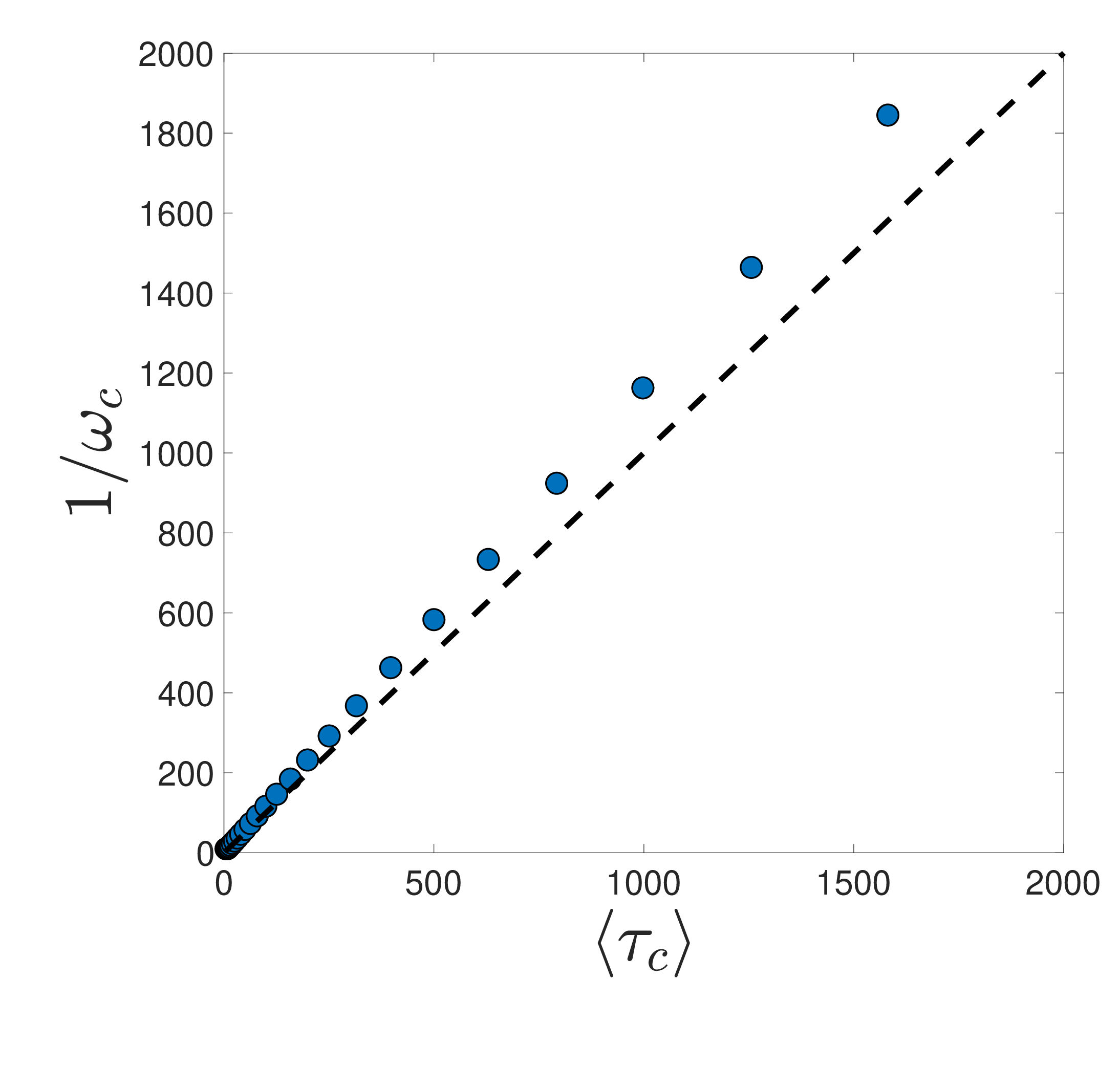}
	\caption{We compare the characteristic relaxation time, defined as $1/\omega_c$, with the average relaxation time over all elements. We find that the characteristic relaxation time can be well approximated by the average relaxation time. In this figure, $D=0.2$, $T=2$, $v=0$. }\label{omegac}
\end{figure}
\bibliography{Lin_Apr_2021}